\documentclass[twocolumn,trackchanges]{aastex63}
\usepackage{xcolor, soul}
\sethlcolor{yellow}
\setstcolor{red}

\newcommand{\code}[1]{\texttt{#1}}
\newcommand{\dStar}{\code{dStar}}
\newcommand*{\EF}{\ensuremath{E_\mathrm{F}}} 
\newcommand*{\kB}{\ensuremath{k_\mathrm{B}}} 
\newcommand*{\Qimp}{\ensuremath{Q_\mathrm{imp}}} 

\newcommand*{\nuclei}[2]{\ensuremath{\mathrm{^{#1}#2}}}

\newcommand*{\oxygen}[1][16]{\nuclei{#1}{O}}
\newcommand*{\neon}[1][20]{\nuclei{#1}{Ne}}
\newcommand*{\magnesium}[1][24]{\nuclei{#1}{Mg}}
\newcommand*{\calcium}[1][40]{\nuclei{#1}{Ca}}

\newcommand*{\chromium}[1][52]{\nuclei{#1}{Cr}}
\newcommand*{\iron}[1][56]{\nuclei{#1}{Fe}}
\newcommand*{\punit}{\ensuremath{\mathrm{dyn\,cm^{-2}}}}
\newcommand*{\rhounit}{\ensuremath{\mathrm{g\,cm^{-3}}}}

\shorttitle{Pycnonuclear Fusion Uncertainties}
\shortauthors{Jain R. et al.}

\watermark{}
\graphicspath{{./}{}}

\begin{document}


\title{Impact of Pycnonuclear Fusion Uncertainties on the Cooling of Accreting Neutron Star Crusts}

\correspondingauthor{R. Jain}
\email{jain@frib.msu.edu}
\author[0000-0001-9859-1512]{R. Jain}
    \affiliation{Facility for Rare Isotope Beams,
    Michigan State University,
    East Lansing MI 48824 USA}
    \affiliation{Department of Physics and Astronomy,
    Michigan State University,
    East Lansing MI 48824 USA}
    \affiliation{Joint Institute for Nuclear Astrophysics-Center for the Evolution of the Elements (JINA-CEE)}
\author[0000-0003-3806-5339]{E. F. Brown}
    \affiliation{Facility for Rare Isotope Beams,
    Michigan State University,
    East Lansing MI 48824 USA}
    \affiliation{Department of Physics and Astronomy,
    Michigan State University,
    East Lansing MI 48824 USA}
    \affiliation{Department of Computational Mathematics, Science and Engineering,
    Michigan State University,
    East Lansing MI 48824 USA}
    \affiliation{Joint Institute for Nuclear Astrophysics-Center for the Evolution of the Elements (JINA-CEE)}
\author[0000-0003-1674-4859]{H. Schatz}
    \affiliation{Facility for Rare Isotope Beams,
    Michigan State University,
    East Lansing MI 48824 USA}
    \affiliation{Department of Physics and Astronomy,
    Michigan State University,
    East Lansing MI 48824 USA}
    \affiliation{Joint Institute for Nuclear Astrophysics-Center for the Evolution of the Elements (JINA-CEE)}
\author{A. V. Afanasjev}
    \affiliation{Department of Physics and Astronomy,
    Mississippi State University,
    Mississippi State, MS 39762, USA}
\author{M. Beard}
    \affiliation{Joint Institute for Nuclear Astrophysics-Center for the Evolution of the Elements (JINA-CEE)}
    \affiliation{Department of Physics,
    University of Notre Dame,
    Notre Dame, IN 46556, USA}
\author{L. R. Gasques}
    \affiliation{Universidade de Sao Paulo, Instituto de Fisica, 
    Rua do Matao, 1371, 05508-090, Sao Paulo, SP, Brazil}
\author{S. S. Gupta}
    \affiliation{Indian Institute of Technology Ropar,
    Rupnagar (Ropar), Punjab 140 001, India}
\author[0000-0003-3161-3283]{G. W. Hitt}
    \affiliation{Department of Physics and Engineering Science,
    Coastal Carolina University,
    P.O. Box 261954, Conway, SC 29528, USA}
\author[0000-0002-9481-9126]{W. R. Hix}
    \affiliation{Physics Division, Oak Ridge National Laboratory,
    P.O. Box 2008, Oak Ridge, TN 37831-6354, USA}
    \affiliation{Department of Physics and Astronomy,
    University of Tennessee Knoxville,
    Knoxville, TN 37996-1200, USA}
\author{R. Lau}
    \affiliation{ HKU SPACE, PO LEUNG KUK, Stanely Ho Community Colllege,
    Hong Kong}
\author{P. M\"{o}ller}
    \affiliation{Joint Institute for Nuclear Astrophysics-Center for the Evolution of the Elements (JINA-CEE)}
\author{W. J. Ong}
    \affiliation{Nuclear and Chemical Sciences Division, 
    Lawerence Livermore National Laboratory,
    Livermore, CA 94551, USA}
    \affiliation{Joint Institute for Nuclear Astrophysics-Center for the Evolution of the Elements (JINA-CEE)}
\author{M. Wiescher}
    \affiliation{Joint Institute for Nuclear Astrophysics-Center for the Evolution of the Elements (JINA-CEE)}
    \affiliation{Department of Physics,
    University of Notre Dame,
    Notre Dame, IN 46556, USA}
\author{Y. Xu}
    \affiliation{Extreme Light Infrastructure - Nuclear Physics (ELI-NP), Horia Hulubei National Institute for R\&D in Physics and Nuclear Engineering (IFIN-HH), 077125 Buchurest-Magurele, Romania}
    
\begin{abstract}
The observation of X-rays during quiescence from transiently accreting neutron stars provides unique clues about the nature of dense matter. This, however, requires extensive modeling of the crusts and matching the results to observations. The pycnonuclear fusion reaction rates implemented in these models are theoretically calculated by extending phenomenological expressions and have large uncertainties spanning many orders of magnitude. We present the first sensitivity studies of these pycnonuclear fusion reactions in realistic network calculations. We also couple the reaction network with the thermal evolution code \code{dStar} to  further study their impact on the neutron star cooling curves in quiescence. Varying the pycnonuclear fusion reaction rates alters the depth at which nuclear heat is deposited although the total heating remains constant. The enhancement of the pycnonuclear fusion reaction rates leads to an overall shallower deposition of nuclear heat. The impurity factors are also altered depending on the type of ashes deposited on the crust. These total changes correspond to a variation of up to 9\,eV in the modeled cooling curves. While this is not sufficient to explain the shallow heat source, it is comparable to the observational uncertainties and can still be important for modeling the neutron star crust. 

\end{abstract}


\section{Introduction} \label{sec:intro}

When neutron stars reside in Low Mass X-ray Binaries (LMXB), they can accrete matter onto their surface from the companion star \citep{Tauri2006}. This accretion can be unstable \citep{Mineshige1993} resulting in transient systems with accretion outbursts and quiescent phases. During accretion, nuclear reactions heat the neutron star crust as the crust is compressed by the infalling matter \citep{Sato1979,Haensel1990,Haensel2008,Gupta2007,Gupta2008,Shternin2007,Meisel2018,Lau2018,Shchechilin2019,Shchechilin2021}. Together with the thermal transport properties, crustal heating determines the thermal structure of the crust. Depending on the depth of the nuclear heat sources, a large fraction of the nuclear heating is directed towards the neutron star core. Crustal heating is therefore also important for modeling the long-term thermal evolution of accreting neutron stars, which can provide constraints on the properties of the neutron star core \citep[e.g.,][]{Brown2018,Ootes2019,Potekhin2019}.

During the quiescence phase, the crust cools on observable timescales of several months to a few years \citep{Rutledge2002,Cackett2006,Shternin2007,Brown2009}. Observations of the cooling neutron star surface therefore probe directly the thermal structure of the crust and its history of nuclear heating. Comparing models of these cooling curves to observations has led to many new insights into the physics of the neutron star crust such as the well-ordered lattice structure in the crust \citep{Cackett2006,Shternin2007,Brown2009}, neutron superfluidity \citep{Shternin2007,Brown2009} and the hints for the existence of nuclear pasta at the crust-core transition \citep{Horowitz2015,Deibel2017}. 

Previous studies \citep{Sato1979,Haensel1990,Haensel2008} have shown that pycnonuclear fusion reactions are the main heat source in crusts of accreting neutron stars. \citet{Yakovlev2006} showed that the theoretical predictions for these rates can be uncertain by 8 to 10 orders of magnitude (See \S~\ref{sec:pycnonuclear}). Here, we investigate for the first time the impact of these uncertainties on the crust temperature profiles and the observable surface cooling during quiescence by varying the pycnonuclear fusion reaction rates in realistic nuclear network calculations.

This is of particular importance in light of previous studies that demonstrated that current estimates of nuclear heating are not sufficient to explain observed quiescent cooling as function of time. Instead, a strong shallow heat source has to be artificially included for models to match observations \citep{Brown2009,Degenaar2015,Page2013,Deibel2015,Turlione2015,Merritt2016,Waterhouse2016,Parikh2019,Chamel2020,Potekhin2021}. The characteristics of this inferred shallow heat source differ from system to system and even from one outburst to the other for the same system \citep{Degenaar2019}, and also depend on the assumed crust composition \citep{Potekhin2021}. Typically models require 0.5--2\,MeV/u (MeV per accreted nucleon) of additional shallow heating, but for some systems up to of the order of 10\,MeV/u \citep{Deibel2015} appears to be needed \citep[for an alternative interpretation, see][]{Page2022A-Hyperburst-in}. The nature and origin of this shallow heat source are unknown. \citet{Horowitz2008} proposed that some of the shallow heating could be due to the nuclear fusion of neutron-rich lighter elements with $A \approx 24\textrm{--}28$ at shallower depths. However, \citet{Lau2018} showed that not enough energy is released by these reactions to account for the required shallow heating strength. \citet{Chamel2020} point out that fusion of $^{12}$C or $^{16}$O with subsequent electron captures could provide up to 1.4\,MeV/u of shallow heat. For realistic abundances, however, the energy release would be considerably smaller. Furthermore, were such light elements present in significant quantities, they would be expected to burn explosively at relatively shallow depths \citep{Cumming2001,Cooper2009Possible-Resona}. Our goal here is to determine whether increased fusion reaction rates, within their uncertainties, may shift some fusion reactions to sufficiently shallow depths to reduce the need for an artificially introduced source of shallow heating. The depth where a fusion reaction occurs not only affects the location, but also the total amount of energy released due to changes in the electron chemical potential that affect subsequent electron capture reactions. 

The next section briefly describes pycnonuclear fusion reaction rates and the corresponding uncertainties associated with their calculations. After that, we discuss how the network is set up and how the cooling curves are calculated. Then we present the results of our sensitivity studies for a variety of initial compositions, particularly focusing on the cases where we see enhanced nuclear heating at shallower depth. We also identify the most important pycnonuclear fusion reactions that contribute to nuclear heating. Subsequently, we discuss the impact of these sensitivities on the predicted cooling curves which can be directly compared to the observations.

\section{Pycnonuclear Fusion Rates and their Uncertainties} \label{sec:pycnonuclear}

Pycnonuclear fusion is density-induced nuclear fusion \citep{Harrison_1964} enabled by the overlapping zero-point energy vibrations of nuclei. The rates used in our reaction network are calculated following the formalism presented in \citet{Yakovlev2006} for a multi-component plasma (MCP).
There are two main contributions to the uncertainties of pycnonuclear fusion reaction rates - nuclear physics uncertainties and plasma physics uncertainties. The nuclear physics uncertainties stem principally from the calculation of the astrophysical S-factors of the reactions. These are calculated using Sao-Paulo potential \citep{Chamon2002,Chamon2007} for describing the nucleus. A universal 9-parameter expression in \citet{Beard2010} was fit to the calculations for 946 reactions involving various isotopes of C, O, Ne and Mg. We use the same parameters for calculating the S-factors of 4844 different reactions from Be to Si in our network. The typical nuclear physics uncertainties are estimated to be two orders of magnitude \citep{Gasques2007,Umar2012}; however, \citet{Ernst2014} determined experimentally the fusion S-factors for $^{10,14,15}$C + $^{12}$C and found excellent agreement with theory. It remains to be seen if this agreement holds for more exotic isotopes and for the fusion of neutron-rich isotopes of heavier elements up to Mg and Si.

In addition to uncertainties in the nuclear physics, plasma physics uncertainties are also present and are even larger than the nuclear uncertainties. They originate from the screening effects due to the Coulomb field of the surrounding plasma particles and uncertain spatial correlations of ions in a multi-component plasma. Overall, the total uncertainties span 8 to 10 orders of magnitude \citep{Yakovlev2006}. Here we use an uncertainty of a factor of $10^{6}$ in either direction to account for the combined effect of various theoretical uncertainties. Calculations are performed with all pycnonuclear fusion reaction rates increased and decreased by a factor of $10^{6}$. Our goal is a ``worst case'' approach with the simultaneous enhancement and reduction of all pycnonuclear fusion rates. Owing to the strong density dependence of the rates, increased rates primarily lead to a shallower heat deposition whereas decreased rates primarily lead to a deeper heat deposition. We therefore expect that changing all rates simultaneously has the largest effect on the depth distribution of the heating, and thus on predictions of the observable cooling curve. We also use variations of a factor of one hundred in both directions to explore the maximum impact of just the estimated nuclear physics uncertainties. 

\section{Model} \label{sec:model}

\subsection{Nuclear Reaction Network Calculations} \label{sec:xnet}

We use the same nuclear reaction network as in \citet{Lau2018} and only summarize it briefly here. The network includes a comprehensive set of nuclear reactions including electron capture, $\beta^{-}$ decay, neutron capture/emission, and nuclear fusion. To calculate the steady state composition of the accreted neutron star crust, we follow the composition changes of an accreted fluid element due to increasing pressure.  In a thin plane-parallel layer, the pressure $P$ can be expressed as $P = \dot{m}gt$, where $\dot{m} = 2.64 \times 10^{4}$ g cm$^{-2}$ s$^{-1}$ is the local accretion rate, $g = 1.85 \times 10^{14}$ cm s$^{-2}$ is the local surface gravity (Newtonian value for a typical neutron star of mass $1.4\,M_{\odot}$ and radius 10\,km), and $t$ is the time since deposition on the surface. Our accretion rate corresponds to 0.3 of the spherical Eddington accretion rate (8.8$\times 10^4$\,g cm$^{-2}$ s$^{-1}$), which is typical for many quasi-persistent transients. The density, $\rho$, is calculated using an equation of state \citep{Gupta2007} that includes degenerate, relativistic electrons, strongly coupled ions, and, at greater depth, free neutrons. In this work, we use pressure to indicate the depth. Temperature is treated as a constant parameter with $T = 0.5$ GK throughout the crust to facilitate comparisons with previous results \citep{Lau2018}. Our analysis does not significantly depend on the choice of temperature as the zero-temperature pycnonuclear fusion reaction rates that are implemented in the network do not depend on temperature.

\begin{figure}[tbp]
\includegraphics[width=\linewidth]{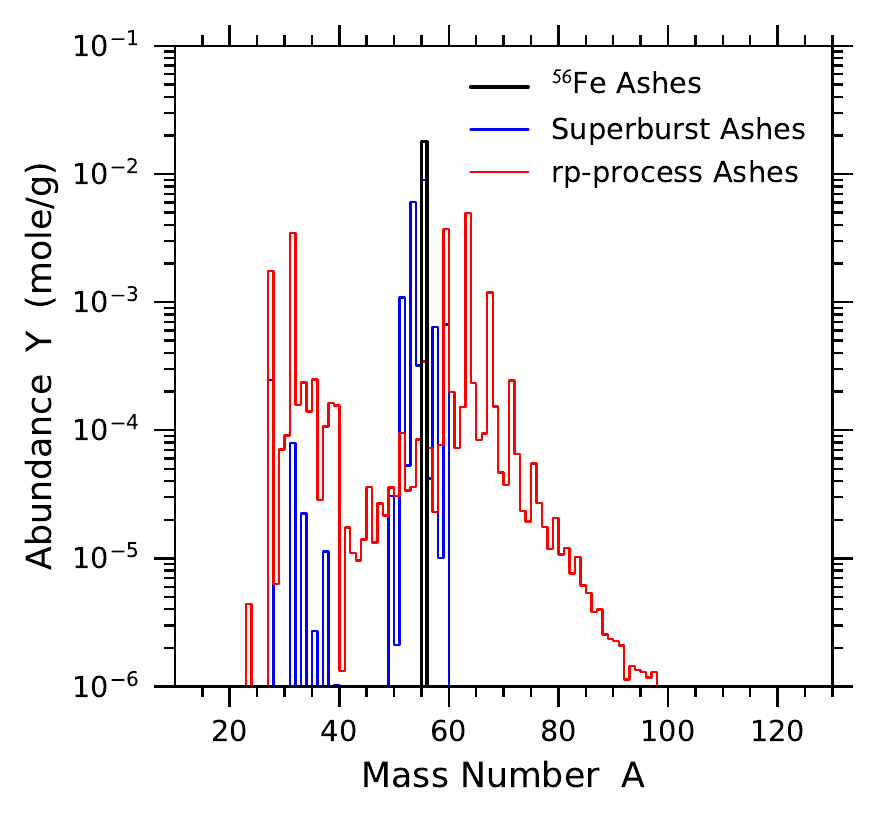}
\caption{Abundances Y (in mole/g) as a function of mass number for the different initial compositions used in this work. The black curve represents a pure  $^{56}$Fe composition. The blue curve corresponds to superburst ashes that have a high abundance of iron peak elements. The red curve corresponds to a typical type-I X-ray burst ashes. They contain a significant fraction of light elements in the mass range $20 < A < 40$ as well as heavier elements with $A > 60$ that are formed by the rp-process.
\label{fig:abd}}
\end{figure}

The initial composition on the surface of the neutron star crust depends on the composition of the accreted matter and the previous thermonuclear burning stages in the atmosphere and the ocean. We first consider an initial composition of $^{56}$Fe for comparison with the previous results by \citet{Haensel1990}. Next, we consider the initial composition of ashes predicted for a superburst powered by explosive carbon burning \citep{Keek2012}. This is appropriate for a system where all accreted matter is processed in repeated superbursts. During superbursts, explosive carbon burning leads to high temperatures and Nuclear Statistical Equilibrium (NSE) is established at relatively low densities and modest neutronization. As a result, the ashes are mainly composed of elements around iron. Finally, we also consider the ashes of hydrogen- and helium-rich X-ray bursts that underwent a rapid proton capture process (rp-process) \citep{Woosley2004,Cyburt2016}. The  rp-process produces nuclei beyond $^{56}$Fe via a sequence of proton captures and $\beta$ decays, and the resulting nuclear ashes contain significant abundances of nuclei with mass number $A > 60$. Figure~\ref{fig:abd} shows the abundances as a function of the mass number for all three initial compositions.

\subsection{Thermal Transport Code} \label{sec:dstar}

We use the thermal transport code \dStar\ \citep{Brown2015} to model the evolution of the thermal structure of the crust during accretion, and the thermal emission from the neutron star surface during quiescence. We model a neutron star of mass $1.62\,M_{\odot}$ and radius 11.2\,km that accretes at a rate of 10$^{17}$ g s$^{-1}$ for 4,383\,days before shutting off into quiescence. The choice of these parameters is motivated by observations of KS~1731-260 \citep{Brown2009,Merritt2016}. For a given crust equation of state, \dStar\ integrates the  relativistic equations of stellar structure \citep{Tolman1939,Oppenheimer1939} to compute the mechanical structure of the crust; \dStar\ then divides the crust into a series of mass shells and integrates the thermal evolution equations \citep{thorne77} using a method-of-lines algorithm \citep{Schiesser1991The-Numerical-M}. 

The energy deposited in each mass shell during accretion is taken from the separate reaction network calculations. No other heating is included in our model. To obtain the heat sources, we run the network at 0.5\,GK without $\beta$-decays to remove any crust Urca cooling \citep{Schatz2013}. Crust Urca cooling is included in \dStar\ directly using the Urca pairs identified in the full network calculation at 0.5\,GK, with their luminosities scaled as $T^{5}$. We find, however, that at our relatively low crust temperatures of up to 0.2\,GK, Urca cooling is negligible. Our approach of decoupling the nuclear reaction network calculation from the cooling model assumes that nuclear heating is independent of temperature. This is justified, as $\kB T \ll \EF$, $\EF$ being the electron Fermi energy. The finite temperature therefore results only in a small shift in depth and a small broadening of the heat deposition by electron capture transitions. In addition, the pycnonuclear fusion reactions rates implemented are independent of temperature. 

Heat transport via electron-phonon, electron-ion, and electron-impurity scattering is included. For electron-ion scattering, the crust composition, in particular the impurity factor, governs the heat transport in the crust \citep{Roggero2016}. The impurity factor is the variance of the atomic numbers of elements present in the crust weighted by their abundances. It is defined as $\Qimp = \Sigma_{i}Y_{i}(Z_{i}-\langle Z \rangle )^{2}/\Sigma_{i}Y_{i}$ where $Y_{i}$ and $Z_{i}$ are the abundance and charge, respectively, of species $i$ and $\langle Z \rangle$ is the average charge number of the composition. The impurity factor in each zone is calculated from the composition as a function of depth obtained from the reaction network calculation. 

The system is first evolved with nuclear heat sources during the accretion phase, establishing a temperature profile in the crust. Heating is then turned off for the subsequent quiescence phase. The calculated evolution of the effective surface temperature with respect to the observer at infinity corresponds to the observable quiescent cooling curve. 

\section{Results} \label{sec:results}

\begin{figure}[tbp]
\includegraphics[width=\linewidth]{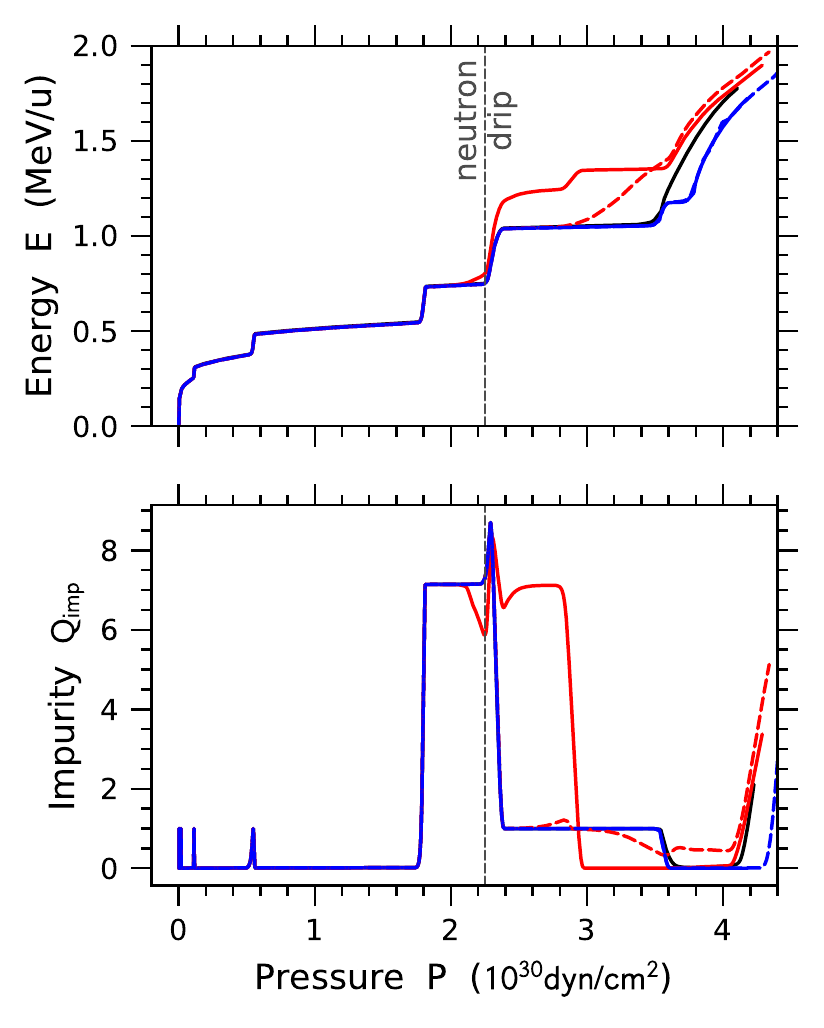}
\caption{The upper panel shows the total integrated energy (in MeV/u) and the lower panel shows the impurity factor ($Q_{\mathrm{imp}}$) as functions of pressure (depth) for our calculations with $^{56}$Fe initial composition. The results are shown for the baseline calculation with nominal reaction network (black solid line), all pycnonuclear fusion reaction rates enhanced by a factor of a million (red solid line), and all rates reduced by a factor of a million (blue solid line). The dashed lines correspond to results with pycnonuclear reaction rates enhanced (red) and reduced (blue) by a factor of a hundred.
\label{fig:fe56}}
\end{figure}

\begin{figure*}[t]
\includegraphics[width=\linewidth]{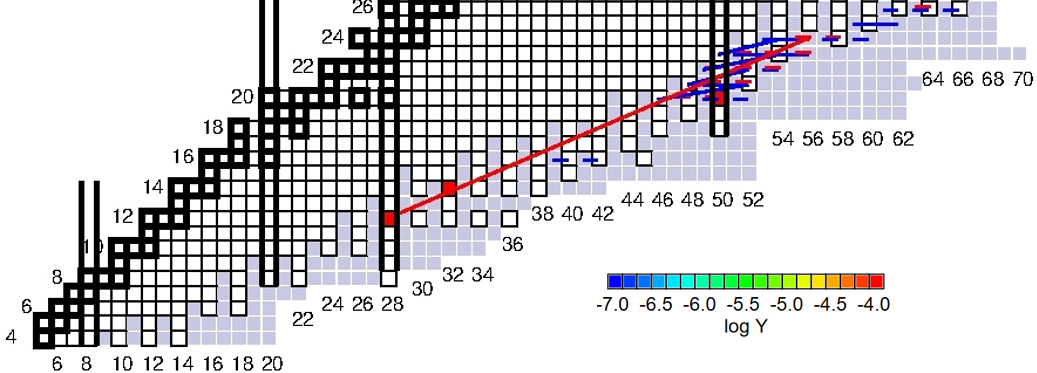}
\caption{Abundances (in mole/g), and reaction flows on the nuclear chart at a depth of $P = 2.5 \times 10^{30}\,\punit$ (or density, $\rho = 1.2 \times 10^{12}\,\rhounit$) for the pycnonuclear fusion reaction rates enhanced 
by a factor  of a million for the scenario of \iron\, ashes initial composition. The rows are labelled on the left with charge number $Z$, and the columns are labelled on the bottom with neutron number $N$. Squares surrounded by thick black lines show stable nuclei and the light gray squares show neutron unbound nuclei. The thick vertical lines correspond to the classical neutron magic numbers. The lines in red are reaction flows towards higher mass number while the lines in blue are for reactions flows towards lower mass number. The buildup of $^{70}$Ca due to enhanced pycnonuclear fusion reaction rates at a relatively shallower depth compared to the default case is shown as discussed in Section \ref{sec:fe56}.
\label{fig:rs1}}
\end{figure*}

\begin{table}
\caption{Important pycnonuclear fusion reactions based on the total integrated flow through the respective channels for $^{56}$Fe initial composition.
\label{table:fe56}}
\begin{tabular}{ c c }
\hline \hline
Reactions & Integrated flow \\
& (mole/g) \\
\hline
$^{40}$Mg + $^{40}$Mg $\rightarrow$\ $^{80}$Cr &  9.142 $\times$ 10$^{-3}$\\
$^{40}$Mg + $^{44}$Mg $\rightarrow$\ $^{84}$Cr &  1.588 $\times$ 10$^{-3}$\\
$^{44}$Mg + $^{44}$Mg $\rightarrow$\ $^{88}$Cr &  7.909 $\times$ 10$^{-4}$\\
$^{38}$Ne + $^{44}$Mg $\rightarrow$\ $^{82}$Ti &  4.090 $\times$ 10$^{-4}$\\
$^{38}$Ne + $^{40}$Mg $\rightarrow$\ $^{78}$Ti &  9.617 $\times$ 10$^{-5}$\\
\hline
\end{tabular}
\end{table}

\subsection[Iron-56]{\iron\ Initial Composition} \label{sec:fe56}

For an initial $^{56}$Fe composition, the upper panel in Figure \ref{fig:fe56} shows the integrated total nuclear heating as a function of depth for the nominal pycnonuclear fusion rates, as well as for fusion rates enhanced and reduced by a factor of a million, respectively. We find that regardless of the rate changes, the same fusion reactions occur (Table \ref{table:fe56}). Therefore, final composition and the total generated heat at the end of the calculation are similar for all cases. However, the changes in the pycnonuclear fusion rates lead to changes in the depth where the respective reactions occur. Consequently, enhanced pycnonuclear fusion reaction rates lead to an overall shallower heat deposition and the reduced reactions rates case shows a deeper heat deposition. A similar, but correspondingly smaller, effect is seen for rate changes of a factor of a hundred. Regardless of the rate changes, the heat deposition from fusion reactions for all calculations with  $^{56}$Fe ashes as initial composition is always beyond $P = 2 \times 10^{30}\,\punit$ and therefore around and beyond the neutron drip.

The rate changes also cause changes in the composition as a function of depth. The lower panel in Figure \ref{fig:fe56} shows that enhanced pycnonuclear fusion makes the crust more impure than the default case. This is because the pycnonuclear fusion of $^{40}$Mg starts earlier at a depth of around $P = 2 \times 10^{30}\,\punit$ (or density, $\rho = 8.8 \times 10^{11}\,\rhounit$), leading to a significant buildup of $^{70}$Ca since the superthreshold electron capture cascades (SEC) \citep{Gupta2008} following pycnonuclear fusion are not fully efficient yet, as shown in Figure \ref{fig:rs1}. This buildup of $^{70}$Ca is absent in the default rates case and is the main reason for the increased impurity of the crust between the depth of $P = 2.4 \times 10^{30}\,\punit$ and $P = 3 \times 10^{30}\,\punit$. Beyond that depth, most of the $^{40}$Mg is depleted due to the enhanced pycnonuclear fusion and the crust impurity becomes lower than the default case. 

The impact of the fusion rate variations on the heating profile is dominated by the $\magnesium[40]+\magnesium[40]$ reaction. Calculations with enhancements of different rate combinations show that the heating profile obtained with just enhancing the $\magnesium[40]+\magnesium[40]$ reaction is the same as the heating profile obtained by enhancing all fusion reactions. Individual enhancements of any of the other rates listed in Table~\ref{table:fe56} do not lead to a significant change in the heat deposition. Enhancing all fusion rates except for the $\magnesium[40]+\magnesium[40]$ reaction does however lead to a modest change in the heat profile. This is due to an enhancement of the fusion reactions involving Si isotopes. What makes $^{40}$Mg so important, nonetheless, is that it has a relatively high electron capture threshold as well as high neutron separation energy for our choice of mass model \citep{Mller2016}. This makes it relatively inert to electron capture as well as neutron capture/emission and the most abundant isotope in the inner neutron star crust.

\begin{figure}[tbp]
\includegraphics[width=\linewidth]{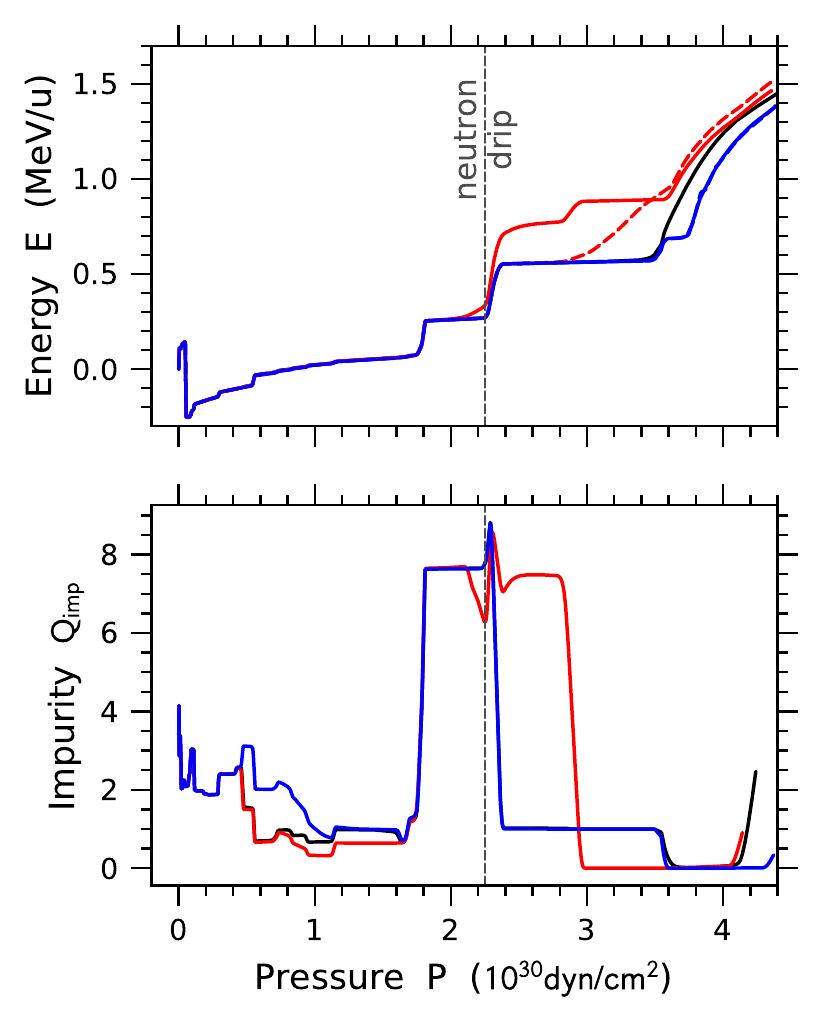}
\caption{Total integrated energy (in MeV/u) and impurity factor ($Q_{\mathrm{imp}}$) as functions of pressure (depth) for an initial composition of superburst ashes. See Figure \ref{fig:fe56} for details.
\label{fig:sup}}
\end{figure}

\subsection{Superburst Ashes Initial Composition} \label{sec:superbursts}
Superburst ashes are mainly composed of elements around $^{56}$Fe (Figure \ref{fig:abd}). Electron capture with neutron emission and neutron capture reactions simplify the composition once neutron drip is reached. Therefore, the composition in the region where pycnonuclear fusion reactions occur is very similar to the pure initial $^{56}$Fe case. As expected, the sensitivity to pycnonuclear fusion reaction rate enhancements and reductions is therefore also very similar (Figure \ref{fig:sup}) to the $^{56}$Fe case. One difference in the integrated energy is the initial cooling via crust Urca cooling \citep{Schatz2013} due to the initial presence of odd-$A$ nuclei in the superburst ashes. This cooling depends sensitively on the chosen crust temperature, but does not impact the heating from pycnonuclear fusion at later times, which is of primary interest here. 

\subsection{Rp-Process Ashes Initial Composition} \label{sec:rpprocess}
\begin{figure}[htbp]
\includegraphics[width=\linewidth]{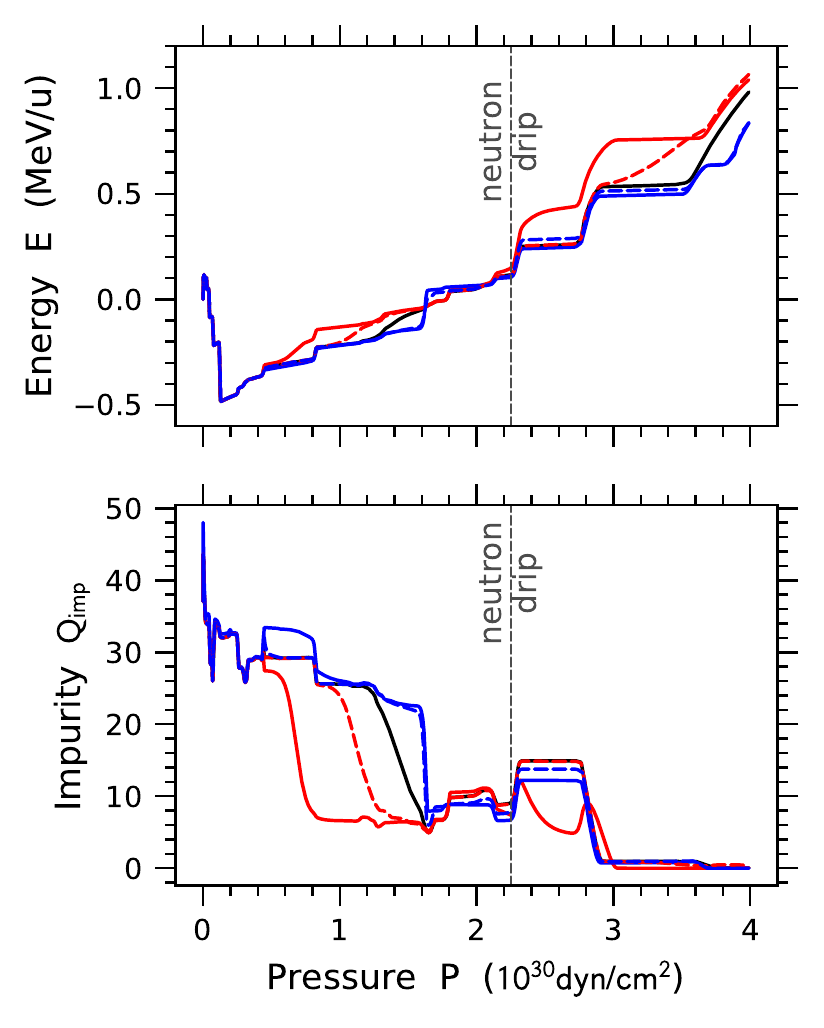}
\caption{Total integrated energy (in MeV/u) and impurity factor ($Q_{\mathrm{imp}}$) as functions of pressure (depth) for an initial composition of rp-process ashes. See Figure \ref{fig:fe56} for details.
\label{fig:rp}}
\end{figure}

Figure \ref{fig:rp} shows the dependence of heating and impurity profiles from changing all pycnonuclear fusion rates by factors of a million and a hundred for an initial composition of rp-process ashes. The main difference to the models with initial $^{56}$Fe or superburst ashes is the presence of fusion reactions at shallower depth \citep{Lau2018}, well before the neutron drip. This is a consequence of the presence of lighter nuclei in the initial composition which are mainly synthesized by secondary burning of residual helium left unburned by the rp-process during an X-ray burst. As a result, increased pycnonuclear fusion reaction rates already significantly increase heating at shallower depths above neutron drip, with heating already occurring at $P = 4 \times 10^{29}\,\punit$ ($\rho = 2.3 \times 10^{11}\,\rhounit$) for the most extreme case of a rate increase by a factor of a million (Figure \ref{fig:rp}). The initial negative heat deposition is again due to Urca cooling as discussed in \S~\ref{sec:superbursts}. The different final total energies for the different cases in Figure~\ref{fig:rp} are an artifact of choosing a fixed pressure to end the calculations. Continued running of the cases with lower fusion rates would eventually lead to the same final composition and similar total energy release.

The reaction sequence for the default case is the same as that discussed in \cite{Lau2018}. Fusion reactions involving isotopes like $^{21}$N, $^{20}$O, and $^{20}$C are initiated around a pressure of $P = 1.8 \times 10^{29}\,\punit$ ($\rho = 1.2 \times 10^{11}\,\rhounit$). However, the heat generation at this stage does not appear to be sensitive to the rates of these reactions. The increased heat deposition at a relatively shallow depth where $P = 4 \times 10^{29}\,\punit$ ($\rho = 2.3 \times 10^{11}\,\rhounit$) for the rates enhanced by a factor of a million is due to fusion of $^{32}$Ne and $^{30}$Ne  producing $^{56}$Ar, $^{62}$Ca, and $^{64}$Ca. The onset of these fusion reactions can also be identified in the corresponding impurity parameter drop of about 20 in Figure \ref{fig:rp}. For the default rates case, the onset of Ne fusion and the associated increase in heating and decrease in impurity occur later, at $P = 1.6 \times 10^{30}\,\punit$ ($\rho = 7.2 \times 10^{11}\,\rhounit$). It is interesting to note that enhanced pycnonuclear fusion increases impurity in the crust for the $^{56}$Fe initial composition (see Figure \ref{fig:rs1}) whereas it decreases impurity for rp-process ashes. This is because for rp-process ashes, the crust is already very impure with the composition spanning a large portion of the nuclear chart. Fusion reactions of the lightest elements present in the rp-process composition occur early and do not lead to fusion cycles in combination with the SEC cascades \citep{Lau2018}. Therefore, these fusion reactions eliminate nuclei at the lower mass end, thereby reducing the spread in atomic number. 

\begin{table}[htb]
\caption{Important pycnonuclear fusion reactions based on the total integrated flow through these channels for rp-process ashes initial composition.
\label{table:rp}}
\centering
\begin{tabular}{cc}
\hline \hline
Reactions & Integrated flow \\
 & (mole/g) \\
\hline
 $\neon[32] + \neon[32] \rightarrow \calcium[64]$ &  $1.691\times10^{-3}$\\
 $\neon[30] + \neon[32] \rightarrow \calcium[62]$ &  $5.288\times10^{-4}$\\
 $\neon[30] + \neon[30] \rightarrow \calcium[60]$ &  $1.079\times10^{-5}$\\
 $\magnesium[40] + \magnesium[40] \rightarrow \chromium[80]$ & $9.846\times10^{-6}$\\
 $\oxygen[24] + \magnesium[40] \rightarrow \calcium[64]$ &  $2.907\times10^{-7}$\\
\hline
\end{tabular}
\end{table}

Around the neutron drip, $^{40}$Mg starts to fuse at $P = 2.2 \times 10^{30}\,\punit$ ($\rho = 9.8 \times 10^{11}\,\rhounit$) for the enhanced rates case and at $P = 2.5 \times 10^{30}\,\punit$ ($\rho = 1.2 \times 10^{12}\,\rhounit$) for the default case. From this point onwards, this case is very similar to the $^{56}$Fe initial composition case. Table \ref{table:rp} lists the most important pycnonuclear fusion reactions in terms of the total integrated flow through their channels. The $^{40}$Mg + $^{40}$Mg reaction with the highest flow in $^{56}$Fe initial composition case is no longer the most important reaction as the fusion of Ne isotopes at shallower depths dominate the total integrated flow.

\section{Impact on Crust Cooling} \label{sec:nscool}

\begin{figure}[tb]
\includegraphics[width=\linewidth]{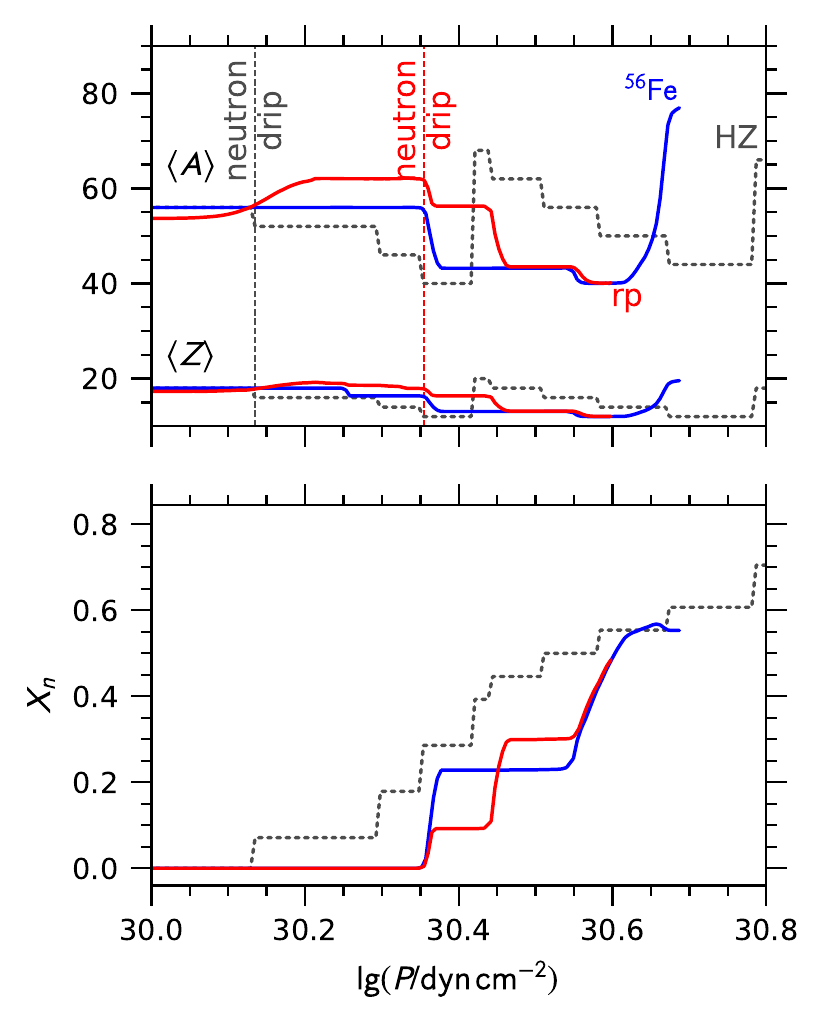}
\caption{(Top) Mean charge and mass number ($\langle Z\rangle$ and $\langle A\rangle$, respectively) for an accreted crust composed of pure \iron\ (red solid line) and one composed of rp-process ashes (blue solid line). For reference, the composition of \citet{Haensel1990} is shown as well (gray dashed line). The bottom panel shows the mass fraction of neutrons for these compositions. In the top panel the vertical lines mark the location of neutron drip; this location is the same for \iron\ and rp-process ashes.
\label{fig:composition}}
\end{figure}

To incorporate the results of our network calculations into models of the thermal evolution of the crust, we first compute the crust equation of state resulting from our calculations.
Figure \ref{fig:composition} shows, in the region around neutron drip, the composition of two sample network calculations: one that starts with an initial composition of pure \iron\ (red curves) and another that starts with a typical rp-process ash (blue curves). Compared with a reference calculation using the default \code{dStar} parameters \citep[][gray curves]{Haensel1990}, our neutron drip occurs at a higher pressure.

\begin{figure}[tb]
\includegraphics[width=\linewidth]{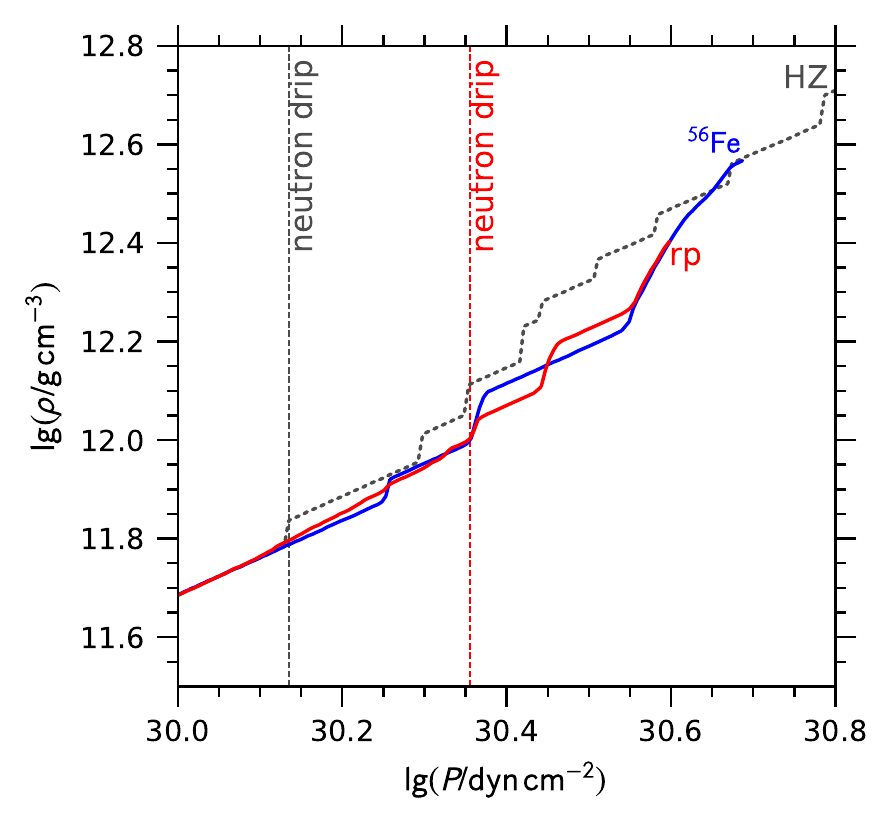}
\caption{Mass density as a function of pressure in the crust for the compositions displayed in Figure \ref{fig:composition}.
\label{fig:eos}}
\end{figure}

The resulting equation of state of these calculations is displayed in Figure ~\ref{fig:eos}. Compared with the reference composition of \citet{Haensel1990}, our calculations have a lower mass density in the vicinity of neutron drip owing to the different nuclear physics, such as the different masses \citep{Mller2016}, used.

As can be noted from Figures~\ref{fig:composition} and \ref{fig:eos}, the network runs do not extend to the crust-core interface but end when the abundances reach the edge of the detailed reaction network. At pressures greater than those reached in the network calculation, we use the composition of \citet{Haensel1990}. To ensure that the transition between the two compositions is numerically well-behaved, we linearly blend between the two compositions over 0.2 dex in $\lg(P)$.

\begin{figure}[t]
\includegraphics[width=\linewidth]{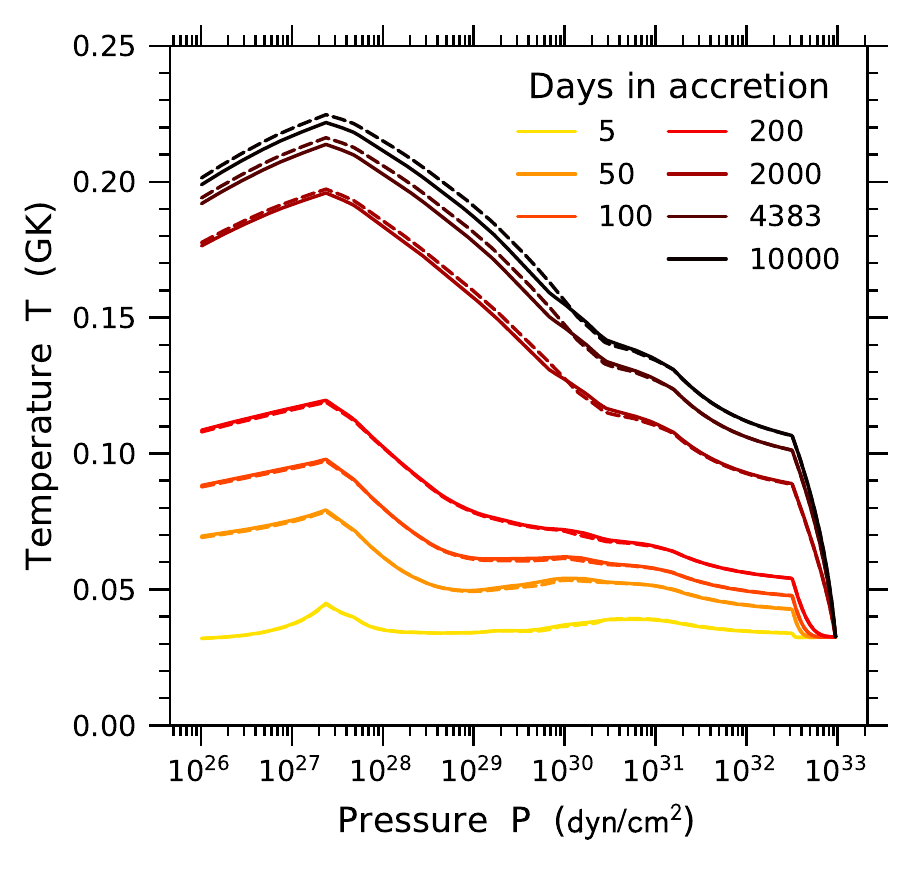}
\caption{The evolution of neutron star temperatures profiles calculated for a crust composed of rp-process ashes, after different number of days in accretion outbursts as calculated with \dStar. Solid lines correspond to calculations with enhanced pycnonuclear fusion reaction rates and dashed lines correspond to calculations with default rates.
\label{fig:tp}}
\end{figure}

The evolution of the surface effective temperature during quiescence was calculated for the various heat source distributions obtained from the reaction network for different assumptions on initial crust composition and pycnonuclear fusion rates.
Figure \ref{fig:tp} shows the evolution of the crust temperature profile for rp-process ashes during an accretion outburst. During accretion, the temperature in the crust rises steadily. While in our model the accretion phase ends at 4,383 days, we show here the evolution for continued accretion out to 10,000 days to demonstrate that at 4,383 days the temperature profile is close to steady state. The kinks at $P = 2 \times 10^{27}$ and $3 \times 10^{32}\,\punit$ correspond to the boundaries of the section of the crust where nuclear heating is implemented. The temperature at the crust-core boundary at $P = 10^{33}\,\punit$ is held fixed as core temperature changes in a single outburst are negligible. Figure \ref{fig:tp} shows that the temperature profiles for enhanced pycnonuclear fusion rates are slightly hotter during initial days of accretion. However, as time progresses, the reduced impurity reduces the thermal conductivity of the crust and the enhanced pycnonuclear fusion reaction rates lead to a cooler crust. Nonetheless, at 4,383 days, the largest temperature difference is only 4.4\,MK or 2\% at $P= 5 \times 10^{29}\,\punit$. 

\begin{figure}[t]
\includegraphics[width=\linewidth]{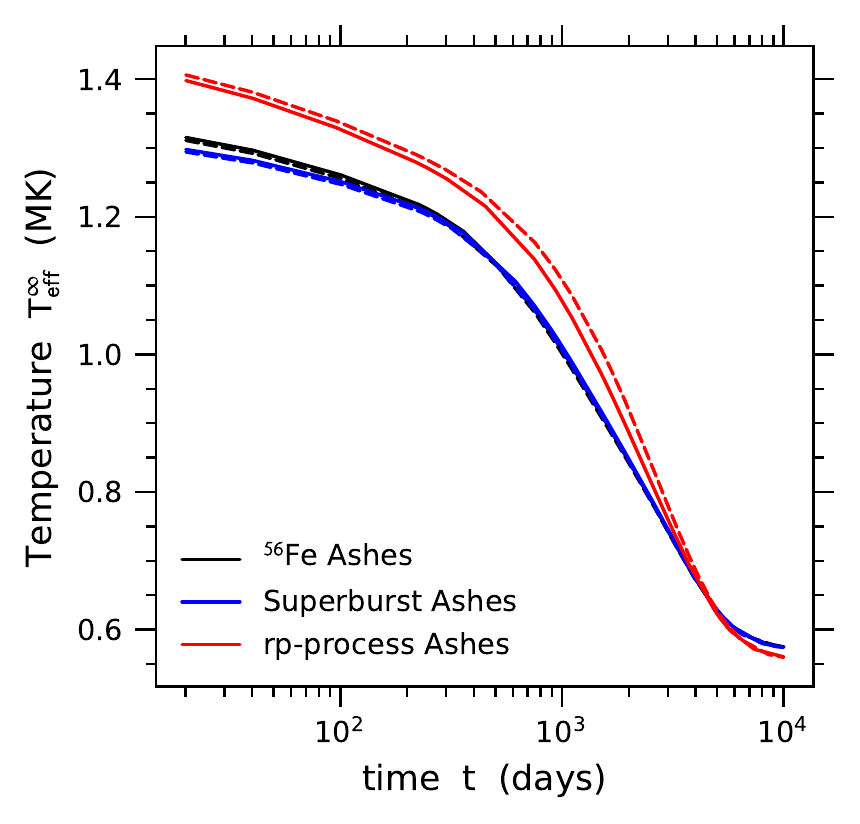}
\caption{Neutron star surface effective temperatures at $\infty$ as a function of the number of days in quiescence calculated with \dStar\. by implementing nuclear heating profiles from reaction network calculations. Solid lines correspond to calculations with enhanced pycnonuclear fusion reaction rates and dashed lines correspond to calculations with default rates. Different colors correspond to different initial compositions explored in this work. 
\label{fig:cc}}
\end{figure}

Cooling curves are then generated by stopping the accretion after 4,383 days (Section \ref{sec:dstar}) and recording the effective surface temperature with respect to the observer at infinity as a function of time. Figure \ref{fig:cc} shows these simulated cooling curves for all three initial compositions and for nominal and enhanced pycnonuclear fusion reaction rates. We find significant differences between the cooling curves with different compositions. The differences with enhancing pycnonuclear fusion reaction rates are the largest at late times for rp-process ashes. Crusts with this composition are also the hottest, as the lighter elements, like Ne, O, and F, in these ashes fuse and deposit heat at shallower depths. As a result, more heat flows to the surface rather than to the core. Even though the heating due to pycnonuclear fusion reactions is similar in crusts with $^{56}$Fe and superburst ashes, the abundances in superburst ashes are distributed among even and odd mass chains affecting the heat released by the electron captures in the outer crust. As a result, the crusts composed of superburst ashes have the lowest surface temperatures.

\section{Conclusions} \label{sec:conclusions}

We provide the results for the first sensitivity studies of pycnonuclear fusion reaction rates in realistic network calculations. When all the pycnonuclear fusion reaction rates are varied up and down by a factor of a million, which corresponds to the estimated uncertainties, the nuclear heat deposition in the inner crust is altered. Owing to the steep density dependence of the pycnonuclear fusion rates, the resulting changes in the predicted cooling curves are relatively small, with temperature changes being at most of the order of 0.1\,MK, or $9\,\mathrm{eV}$ in photon energy. Even though the uncertainties in X-ray observations vary greatly with respect to the source, the telescope, and the time in quiescence, some systems have been observed at this and better levels of statistical uncertainty \citep{Cackett2010,Cackett2013,Fridriksson2011,Degenaar2014}. With newer surveys planned and more sources observed, these differences in the cooling curves can still play an important role in precision modeling of the crust to match the observations. 

When the reaction rates are enhanced, we find an overall shallower heat deposition whereas the reduced reaction rates lead to a deeper heat deposition in general. 
The shallowest pycnonuclear fusion in all the models considered occurs at the depth of $P = 4 \times 10^{29}\,\punit$ ($\rho = 2.3 \times 10^{11}\,\rhounit$) for the rp-process ashes initial composition case with the rates enhanced by a factor of a million. However, the additional shallow heating required to explain observations is typically inferred to occur at a much shallower depth of $P \sim 10^{27}\,\punit$ ($\rho \sim 2 \times 10^{9}\,\rhounit$). We therefore conclude that the uncertainties in pycnonuclear fusion reaction rates are unlikely to explain the shallow heating phenomenon, and will not affect its inferred strength and depth significantly.

During preparation of this work, it was suggested that the diffusion of neutrons may lead to accreted inner crusts that are closer in composition to a cold catalyzed crust with the transition occurring near, but not exactly at, traditional neutron drip \citep{Gusakov2020,Gusakov2021}. In this picture, our results for $P < 2.25 \times 10^{30}\,\punit$ would not be affected. There would be no pycnonuclear fusion beyond $P \approx 2.25 \times 10^{30}\,\punit$ but the energy would be released at the transition pressure. Note that the specific nuclear reaction pathways for this energy release have not been identified. The overall heat release in the crust would only be 10--30\% of the heating reported in previous work \citep[see also][]{Shchechilin2021}.

\acknowledgments

We thank the contributions from D. Yakovlev and P. Shternin as well as the discussion within JINA-CEE crust working group. This work was supported by the National Science Foundation under Award Nos. PHY-1430152 (JINA Center for the Evolution of the Elements), OISE-1927130 (IReNA), PHY-1913554, and PHY-2209429, and by the U.S. Department of Energy, Office of Science, Office of Nuclear Physics under Award No. DE-SC0013037. E.F.B. acknowledges support under grant 80NSSC20K0503 from NASA. This was work also performed under the auspices of the U.S. Department of Energy by Lawrence Livermore National Laboratory under Contract DE-AC52-07NA27344.

\bibliography{main}{}

\begin{thebibliography}{}
\expandafter\ifx\csname natexlab\endcsname\relax\def\natexlab#1{#1}\fi
\providecommand{\url}[1]{\href{#1}{#1}}
\providecommand{\dodoi}[1]{doi:~\href{http://doi.org/#1}{\nolinkurl{#1}}}
\providecommand{\doeprint}[1]{\href{http://ascl.net/#1}{\nolinkurl{http://ascl.net/#1}}}
\providecommand{\doarXiv}[1]{\href{https://arxiv.org/abs/#1}{\nolinkurl{https://arxiv.org/abs/#1}}}

\bibitem[{Beard {et~al.}(2010)Beard, Afanasjev, Chamon, Gasques, Wiescher, \&
  Yakovlev}]{Beard2010}
Beard, M., Afanasjev, A.~V., Chamon, L.~C., {et~al.} 2010, Atomic Data and
  Nuclear Data Tables, 96, 541, \dodoi{10.1016/j.adt.2010.02.005}

\bibitem[{{Brown}(2015)}]{Brown2015}
{Brown}, E.~F. 2015, {dStar: Neutron star thermal evolution code}.
\newblock \doeprint{1505.034}

\bibitem[{{Brown} \& {Cumming}(2009)}]{Brown2009}
{Brown}, E.~F., \& {Cumming}, A. 2009, \apj, 698, 1020,
  \dodoi{10.1088/0004-637X/698/2/1020}

\bibitem[{{Brown} {et~al.}(2018){Brown}, {Cumming}, {Fattoyev}, {Horowitz},
  {Page}, \& {Reddy}}]{Brown2018}
{Brown}, E.~F., {Cumming}, A., {Fattoyev}, F.~J., {et~al.} 2018, \prl, 120,
  182701, \dodoi{10.1103/PhysRevLett.120.182701}

\bibitem[{Cackett {et~al.}(2013)Cackett, Brown, Cumming, Degenaar, Fridriksson,
  Homan, Miller, \& Wijnands}]{Cackett2013}
Cackett, E.~M., Brown, E.~F., Cumming, A., {et~al.} 2013, The Astrophysical
  Journal, 774, 131, \dodoi{10.1088/0004-637x/774/2/131}

\bibitem[{Cackett {et~al.}(2010)Cackett, Brown, Cumming, Degenaar, Miller, \&
  Wijnands}]{Cackett2010}
---. 2010, The Astrophysical Journal, 722, L137,
  \dodoi{10.1088/2041-8205/722/2/l137}

\bibitem[{{Cackett} {et~al.}(2006){Cackett}, {Wijnands}, {Linares}, {Miller},
  {Homan}, \& {Lewin}}]{Cackett2006}
{Cackett}, E.~M., {Wijnands}, R., {Linares}, M., {et~al.} 2006, \mnras, 372,
  479, \dodoi{10.1111/j.1365-2966.2006.10895.x}

\bibitem[{Carnelli {et~al.}(2014)Carnelli, Almaraz-Calderon, Rehm, Albers,
  Alcorta, Bertone, Digiovine, Esbensen, Niello, Henderson, Jiang, Lai, Marley,
  Nusair, Palchan-Hazan, Pardo, Paul, \& Ugalde}]{Ernst2014}
Carnelli, P. F.~F., Almaraz-Calderon, S., Rehm, K.~E., {et~al.} 2014, Phys.
  Rev. Lett., 112, 192701, \dodoi{10.1103/PhysRevLett.112.192701}

\bibitem[{{Chamel} {et~al.}(2020){Chamel}, {Fantina}, {Zdunik}, \&
  {Haensel}}]{Chamel2020}
{Chamel}, N., {Fantina}, A.~F., {Zdunik}, J.~L., \& {Haensel}, P. 2020, \prc,
  102, 015804, \dodoi{10.1103/PhysRevC.102.015804}

\bibitem[{Chamon(2007)}]{Chamon2007}
Chamon, L. 2007, Nuclear Physics A, 787, 198,
  \dodoi{10.1016/j.nuclphysa.2006.12.032}

\bibitem[{Chamon {et~al.}(2002)Chamon, Carlson, Gasques, Pereira, Conti,
  Alvarez, Hussein, Ribeiro, Rossi, \& Silva}]{Chamon2002}
Chamon, L.~C., Carlson, B.~V., Gasques, L.~R., {et~al.} 2002, Physical Review
  C, 66, \dodoi{10.1103/physrevc.66.014610}

\bibitem[{Cooper {et~al.}(2009)Cooper, Steiner, \&
  Brown}]{Cooper2009Possible-Resona}
Cooper, R.~L., Steiner, A.~W., \& Brown, E.~F. 2009, \apj, 702, 660

\bibitem[{{Cumming} \& {Bildsten}(2001)}]{Cumming2001}
{Cumming}, A., \& {Bildsten}, L. 2001, \apjl, 559, L127, \dodoi{10.1086/323937}

\bibitem[{{Cyburt} {et~al.}(2016){Cyburt}, {Amthor}, {Heger}, {Johnson},
  {Keek}, {Meisel}, {Schatz}, \& {Smith}}]{Cyburt2016}
{Cyburt}, R.~H., {Amthor}, A.~M., {Heger}, A., {et~al.} 2016, \apj, 830, 55,
  \dodoi{10.3847/0004-637X/830/2/55}

\bibitem[{Degenaar {et~al.}(2014)Degenaar, Medin, Cumming, Wijnands, Wolff,
  Cackett, Miller, Jonker, Homan, \& Brown}]{Degenaar2014}
Degenaar, N., Medin, Z., Cumming, A., {et~al.} 2014, The Astrophysical Journal,
  791, 47, \dodoi{10.1088/0004-637x/791/1/47}

\bibitem[{{Degenaar} {et~al.}(2015){Degenaar}, {Wijnands}, {Bahramian},
  {Sivakoff}, {Heinke}, {Brown}, {Fridriksson}, {Homan}, {Cackett}, {Cumming},
  {Miller}, {Altamirano}, \& {Pooley}}]{Degenaar2015}
{Degenaar}, N., {Wijnands}, R., {Bahramian}, A., {et~al.} 2015, \mnras, 451,
  2071, \dodoi{10.1093/mnras/stv1054}

\bibitem[{{Degenaar} {et~al.}(2019){Degenaar}, {Ootes}, {Page}, {Wijnands},
  {Parikh}, {Homan}, {Cackett}, {Miller}, {Altamirano}, \&
  {Linares}}]{Degenaar2019}
{Degenaar}, N., {Ootes}, L.~S., {Page}, D., {et~al.} 2019, \mnras, 488, 4477,
  \dodoi{10.1093/mnras/stz1963}

\bibitem[{{Deibel} {et~al.}(2015){Deibel}, {Cumming}, {Brown}, \&
  {Page}}]{Deibel2015}
{Deibel}, A., {Cumming}, A., {Brown}, E.~F., \& {Page}, D. 2015, \apjl, 809,
  L31, \dodoi{10.1088/2041-8205/809/2/L31}

\bibitem[{{Deibel} {et~al.}(2017){Deibel}, {Cumming}, {Brown}, \&
  {Reddy}}]{Deibel2017}
{Deibel}, A., {Cumming}, A., {Brown}, E.~F., \& {Reddy}, S. 2017, \apj, 839,
  95, \dodoi{10.3847/1538-4357/aa6a19}

\bibitem[{Fridriksson {et~al.}(2011)Fridriksson, Homan, Wijnands, Cackett,
  Altamirano, Degenaar, Brown, M{\'{e}}ndez, \& Belloni}]{Fridriksson2011}
Fridriksson, J.~K., Homan, J., Wijnands, R., {et~al.} 2011, The Astrophysical
  Journal, 736, 162, \dodoi{10.1088/0004-637x/736/2/162}

\bibitem[{Gasques {et~al.}(2007)Gasques, Afanasjev, Beard, Lubian, Neff,
  Wiescher, \& Yakovlev}]{Gasques2007}
Gasques, L.~R., Afanasjev, A.~V., Beard, M., {et~al.} 2007, Physical Review C,
  76, \dodoi{10.1103/physrevc.76.045802}

\bibitem[{{Gupta} {et~al.}(2007){Gupta}, {Brown}, {Schatz}, {M{\"o}ller}, \&
  {Kratz}}]{Gupta2007}
{Gupta}, S., {Brown}, E.~F., {Schatz}, H., {M{\"o}ller}, P., \& {Kratz}, K.-L.
  2007, \apj, 662, 1188, \dodoi{10.1086/517869}

\bibitem[{Gupta {et~al.}(2008)Gupta, Kawano, \& M\"{o}ller}]{Gupta2008}
Gupta, S.~S., Kawano, T., \& M\"{o}ller, P. 2008, Physical Review Letters, 101,
  \dodoi{10.1103/physrevlett.101.231101}

\bibitem[{{Gusakov} \& {Chugunov}(2020)}]{Gusakov2020}
{Gusakov}, M.~E., \& {Chugunov}, A.~I. 2020, \prl, 124, 191101,
  \dodoi{10.1103/PhysRevLett.124.191101}

\bibitem[{{Gusakov} \& {Chugunov}(2021)}]{Gusakov2021}
---. 2021, \prd, 103, L101301, \dodoi{10.1103/PhysRevD.103.L101301}

\bibitem[{{Haensel} \& {Zdunik}(1990)}]{Haensel1990}
{Haensel}, P., \& {Zdunik}, J.~L. 1990, \aap, 227, 431

\bibitem[{Haensel \& Zdunik(2008)}]{Haensel2008}
Haensel, P., \& Zdunik, J.~L. 2008, Astronomy {\&} Astrophysics, 480, 459,
  \dodoi{10.1051/0004-6361:20078578}

\bibitem[{Harrison(1964)}]{Harrison_1964}
Harrison, E.~R. 1964, Proceedings of the Physical Society, 84, 213,
  \dodoi{10.1088/0370-1328/84/2/304}

\bibitem[{{Horowitz} {et~al.}(2015){Horowitz}, {Berry}, {Briggs}, {Caplan},
  {Cumming}, \& {Schneider}}]{Horowitz2015}
{Horowitz}, C.~J., {Berry}, D.~K., {Briggs}, C.~M., {et~al.} 2015, \prl, 114,
  031102, \dodoi{10.1103/PhysRevLett.114.031102}

\bibitem[{{Horowitz} {et~al.}(2008){Horowitz}, {Dussan}, \&
  {Berry}}]{Horowitz2008}
{Horowitz}, C.~J., {Dussan}, H., \& {Berry}, D.~K. 2008, \prc, 77, 045807,
  \dodoi{10.1103/PhysRevC.77.045807}

\bibitem[{{Keek} {et~al.}(2012){Keek}, {Heger}, \& {in't Zand}}]{Keek2012}
{Keek}, L., {Heger}, A., \& {in't Zand}, J.~J.~M. 2012, \apj, 752, 150,
  \dodoi{10.1088/0004-637X/752/2/150}

\bibitem[{Lau {et~al.}(2018)Lau, Beard, Gupta, Schatz, Afanasjev, Brown,
  Deibel, Gasques, Hitt, Hix, Keek, M\"{o}ller, Shternin, Steiner, Wiescher, \&
  Xu}]{Lau2018}
Lau, R., Beard, M., Gupta, S.~S., {et~al.} 2018, The Astrophysical Journal,
  859, 62, \dodoi{10.3847/1538-4357/aabfe0}

\bibitem[{Meisel {et~al.}(2018)Meisel, Deibel, Keek, Shternin, \&
  Elfritz}]{Meisel2018}
Meisel, Z., Deibel, A., Keek, L., Shternin, P., \& Elfritz, J. 2018, Journal of
  Physics G: Nuclear and Particle Physics, 45, 093001,
  \dodoi{10.1088/1361-6471/aad171}

\bibitem[{{Merritt} {et~al.}(2016){Merritt}, {Cackett}, {Brown}, {Page},
  {Cumming}, {Degenaar}, {Deibel}, {Homan}, {Miller}, \& {Wijnand
  s}}]{Merritt2016}
{Merritt}, R.~L., {Cackett}, E.~M., {Brown}, E.~F., {et~al.} 2016, \apj, 833,
  186, \dodoi{10.3847/1538-4357/833/2/186}

\bibitem[{{Mineshige}(1993)}]{Mineshige1993}
{Mineshige}, S. 1993, \apss, 210, 83, \dodoi{10.1007/BF00657876}

\bibitem[{M\"{o}ller {et~al.}(2016)M\"{o}ller, Sierk, Ichikawa, \&
  Sagawa}]{Mller2016}
M\"{o}ller, P., Sierk, A., Ichikawa, T., \& Sagawa, H. 2016, At. Data Nucl.
  Data Tables, 109-110, 1, \dodoi{10.1016/j.adt.2015.10.002}

\bibitem[{{Ootes} {et~al.}(2019){Ootes}, {Wijnands}, \& {Page}}]{Ootes2019}
{Ootes}, L.~S., {Wijnands}, R., \& {Page}, D. 2019, \aap, 630, A95,
  \dodoi{10.1051/0004-6361/201936035}

\bibitem[{Oppenheimer \& Volkoff(1939)}]{Oppenheimer1939}
Oppenheimer, J.~R., \& Volkoff, G.~M. 1939, Phys. Rev., 55, 374,
  \dodoi{10.1103/PhysRev.55.374}

\bibitem[{{Page} {et~al.}(2022){Page}, {Homan}, {Nava-Callejas}, {Cavecchi},
  {Beznogov}, {Degenaar}, {Wijnands}, \& {Parikh}}]{Page2022A-Hyperburst-in}
{Page}, D., {Homan}, J., {Nava-Callejas}, M., {et~al.} 2022, arXiv e-prints,
  arXiv:2202.03962.
\newblock \doarXiv{2202.03962}

\bibitem[{{Page} \& {Reddy}(2013)}]{Page2013}
{Page}, D., \& {Reddy}, S. 2013, \prl, 111, 241102,
  \dodoi{10.1103/PhysRevLett.111.241102}

\bibitem[{{Parikh} {et~al.}(2019){Parikh}, {Wijnands}, {Ootes}, {Page},
  {Degenaar}, {Bahramian}, {Brown}, {Cackett}, {Cumming}, {Heinke}, {Homan},
  {Rouco Escorial}, \& {Wijngaarden}}]{Parikh2019}
{Parikh}, A.~S., {Wijnands}, R., {Ootes}, L.~S., {et~al.} 2019, \aap, 624, A84,
  \dodoi{10.1051/0004-6361/201834412}

\bibitem[{{Potekhin} \& {Chabrier}(2021)}]{Potekhin2021}
{Potekhin}, A.~Y., \& {Chabrier}, G. 2021, \aap, 645, A102,
  \dodoi{10.1051/0004-6361/202039006}

\bibitem[{{Potekhin} {et~al.}(2019){Potekhin}, {Chugunov}, \&
  {Chabrier}}]{Potekhin2019}
{Potekhin}, A.~Y., {Chugunov}, A.~I., \& {Chabrier}, G. 2019, \aap, 629, A88,
  \dodoi{10.1051/0004-6361/201936003}

\bibitem[{Roggero \& Reddy(2016)}]{Roggero2016}
Roggero, A., \& Reddy, S. 2016, Physical Review C, 94,
  \dodoi{10.1103/physrevc.94.015803}

\bibitem[{{Rutledge} {et~al.}(2002){Rutledge}, {Bildsten}, {Brown}, {Pavlov},
  {Zavlin}, \& {Ushomirsky}}]{Rutledge2002}
{Rutledge}, R.~E., {Bildsten}, L., {Brown}, E.~F., {et~al.} 2002, \apj, 580,
  413, \dodoi{10.1086/342745}

\bibitem[{{Sato}(1979)}]{Sato1979}
{Sato}, K. 1979, Progress of Theoretical Physics, 62, 957,
  \dodoi{10.1143/PTP.62.957}

\bibitem[{Schatz {et~al.}(2013)Schatz, Gupta, M\"{o}ller, Beard, Brown, Deibel,
  Gasques, Hix, Keek, Lau, Steiner, \& Wiescher}]{Schatz2013}
Schatz, H., Gupta, S., M\"{o}ller, P., {et~al.} 2013, Nature, 505, 62,
  \dodoi{10.1038/nature12757}

\bibitem[{Schiesser(1991)}]{Schiesser1991The-Numerical-M}
Schiesser, W.~E. 1991, The Numerical Method of Lines: Integration of Partial
  Differential Equations (Academic Press)

\bibitem[{{Shchechilin} \& {Chugunov}(2019)}]{Shchechilin2019}
{Shchechilin}, N.~N., \& {Chugunov}, A.~I. 2019, \mnras, 490, 3454,
  \dodoi{10.1093/mnras/stz2838}

\bibitem[{{Shchechilin} {et~al.}(2021){Shchechilin}, {Gusakov}, \&
  {Chugunov}}]{Shchechilin2021}
{Shchechilin}, N.~N., {Gusakov}, M.~E., \& {Chugunov}, A.~I. 2021, arXiv
  e-prints, arXiv:2105.01991.
\newblock \doarXiv{2105.01991}

\bibitem[{{Shternin} {et~al.}(2007){Shternin}, {Yakovlev}, {Haensel}, \&
  {Potekhin}}]{Shternin2007}
{Shternin}, P.~S., {Yakovlev}, D.~G., {Haensel}, P., \& {Potekhin}, A.~Y. 2007,
  \mnras, 382, L43, \dodoi{10.1111/j.1745-3933.2007.00386.x}

\bibitem[{{Tauris} \& {van den Heuvel}(2006)}]{Tauri2006}
{Tauris}, T.~M., \& {van den Heuvel}, E.~P.~J. 2006, in Cambridge Astrophysics
  Series, Vol.~39, Compact Stellar X-ray Sources, ed. W.~Lewin \& M.~{van der
  Klis} (Cambridge University Press), 623--665

\bibitem[{Thorne(1977)}]{thorne77}
Thorne, K.~S. 1977, \apj, 212, 825

\bibitem[{Tolman(1939)}]{Tolman1939}
Tolman, R.~C. 1939, Phys. Rev., 55, 364, \dodoi{10.1103/PhysRev.55.364}

\bibitem[{{Turlione} {et~al.}(2015){Turlione}, {Aguilera}, \&
  {Pons}}]{Turlione2015}
{Turlione}, A., {Aguilera}, D.~N., \& {Pons}, J.~A. 2015, \aap, 577, A5,
  \dodoi{10.1051/0004-6361/201322690}

\bibitem[{Umar {et~al.}(2012)Umar, Oberacker, \& Horowitz}]{Umar2012}
Umar, A.~S., Oberacker, V.~E., \& Horowitz, C.~J. 2012, Physical Review C, 85,
  \dodoi{10.1103/physrevc.85.055801}

\bibitem[{{Waterhouse} {et~al.}(2016){Waterhouse}, {Degenaar}, {Wijnands},
  {Brown}, {Miller}, {Altamirano}, \& {Linares}}]{Waterhouse2016}
{Waterhouse}, A.~C., {Degenaar}, N., {Wijnands}, R., {et~al.} 2016, \mnras,
  456, 4001, \dodoi{10.1093/mnras/stv2959}

\bibitem[{{Woosley} {et~al.}(2004){Woosley}, {Heger}, {Cumming}, {Hoffman},
  {Pruet}, {Rauscher}, {Fisker}, {Schatz}, {Brown}, \&
  {Wiescher}}]{Woosley2004}
{Woosley}, S.~E., {Heger}, A., {Cumming}, A., {et~al.} 2004, \apjs, 151, 75,
  \dodoi{10.1086/381533}

\bibitem[{Yakovlev {et~al.}(2006)Yakovlev, Gasques, Afanasjev, Beard, \&
  Wiescher}]{Yakovlev2006}
Yakovlev, D.~G., Gasques, L.~R., Afanasjev, A.~V., Beard, M., \& Wiescher, M.
  2006, Phys. Rev. C, 74, 035803, \dodoi{10.1103/PhysRevC.74.035803}

\end{thebibliography}
\bibliographystyle{aasjournal}

\listofchanges
\end{document}